# GRSUS: Management Of Health Resources, A Study From The Perspective Of GM/MS 1631/2015 Ordinance In The State Of Pará

# GRSUS: Gerenciamento De Recursos De Saúde, Um Estudo Sob A Ótica Da Portaria GM/MS 1631/2015 No Estado do Pará

Paulo Sérgio Viegas Bernardino da Silva, 0000-0002-8159-4888, (Universidade de Coimbra, Coimbra, Portugal) – pauloviegas93@gmail.com

Lucas Vinícius Araújo Caldas, 0000-0002-9866-5873, (Universidade Federal do Pará, Pará, Brasil) – lvinicius123@gmail.com

Antônio Fernando Lavareda Jacob Junior, 0000-0002-9415-7265, (Universidade Estadual do Maranhão, Maranhão, Brasil) – antoniojunior@professor.uema.br

Fábio Manoel França Lobato, 0000-0002-6282-0368, (Universidade do Oeste do Pará, Pará, Brasil) – fabio.lobato@ufopa.edu.br

Abstract: Investments in public health had an increase of about R$ 20 bi in recent years. Even with the dynamism of the Unique Health System (SUS), only after 13 years the criteria and parameters for the planning and programming of health services have been updated. The calculations for health resources division are complex due to the nature of the SUS administrative organization, which has three administrative levels. Despite providing the criteria and parameters for the calculations, it was not provided any information system that would automate this process and provide reliable information for decision making. In order to fill such gap, this paper presents a system for health resource management from the perspective of GM/MS 1631/2015 ordinance. The tool has been validated using as case studies two municipalities in the interior of the state of Pará. The results were promising, with latent market potential, being possible to simulate various scenarios for a medium and long-term predictions.

Keywords: Health Information System, SUS, Public Policies, Ordinance 1631/2015

Resumo: Os investimentos em saúde pública tiveram um aumento de cerca de 20 bilhões de reais nos últimos anos. Mesmo com a dinamicidade do Sistema Único de Saúde (SUS), apenas após 13 anos os critérios e parâmetros para o planejamento e programação de ações e serviços de saúde foram atualizados. Os cálculos para divisão de recursos de saúde são complexos devido a natureza administrativa do SUS, o qual possui três níveis administrativos. Apesar de fornecer os critérios e parâmetros para os cálculos, não foi disponibilizado nenhum sistema de informação que automatizasse tal processo e fornecesse suporte à decisão. Visando preencher tal lacuna, este trabalho apresenta um sistema de gestão de recursos de saúde sob a ótica da Portaria GM/MS 1631/2015. A ferramenta foi validada utilizando como estudos de caso dois municípios do interior do estado do Pará. Os resultados mostraram-se promissores, com potencial mercadológico latente, sendo possível simular diversos cenários para médio e longo prazo.

Palavras chave: Sistema de Informação aplicados à Saúde, SUS, Políticas Públicas, Portaria 1631/2015

**1. Introdução**

O Brasil foi um dos primeiros e poucos países fora da Organização para a Cooperação e Desenvolvimento Econômico a prever em sua legislação o acesso universal aos serviços de saúde, passando a ser tratado como um dever do Estado. De acordo com o *The World Bank* estima-se que cerca de 70% da população brasileira utiliza-se dos serviços do Sistema Único de Saúde (SUS) e que o Brasil investe cerca de 150 dólares por pessoa na área da saúde.

Devido às diversas realidades regionais de um país com dimensões continentais e à estrutura administrativa do SUS, a gerência da distribuição de recursos é uma tarefa extremamente complexa. Aliado à isso, ainda temos a dinamicidade da saúde, com novos exames, procedimentos e padrões sendo criados e atualizados continuamente. Neste contexto, em 2015 - após 13 anos, os critérios e parâmetros para o planejamento e programação de ações e serviços de saúde no âmbito do SUS foram atualizados pela Portaria 1631, de 1º de outubro de 2015, do Gabinete do Ministério (GM) do Ministério da Saúde (MS), conhecida por Portaria GM/MS 1631/2015. Esta portaria veio substituir a Portaria nº 1.101/GM/MS de 12 de junho de 2002.

A Portaria GM/MS 1631/2015 veio subsidiar o cálculo das estimativas de necessidades da população, definindo critérios que orientem a programação de recursos destinados a investimentos que visem reduzir as desigualdades na oferta de ações e serviços de saúde e garantir a integralidade da atenção à saúde; e foi resultado das pactuações das seguintes comissões de Intergestores: i) regionais, à nível municipal; ii) bipartite, à nível Estadual-Municipal; e iii) tripartite, à nível Federal-Estadual-Municipal.

Nesta portaria, são estabelecidos, por exemplo, parâmetros de cobertura das consultas médicas de urgência especializadas, cirurgias ambulatoriais especializadas, procedimentos traumato-ortopédicos, patologia clínica, anatomopatologia e citopatologia, radiodiagnóstico, exames ultrassonográficos, diagnose, fisioterapia, terapias especializadas, próteses e órteses, ressonância magnética, medicina nuclear *in vitro*, tomografia computadorizada, hemoterapia, entre outros (SOUZA E LAGES, 2012, p. 178).

Além dos itens supramencionados, há muitos outros, o que torna a análise de distribuição de recursos uma tarefa trabalhosa, sendo inviável realizá-la manualmente. Mesmo assim, o Ministério da Saúde não disponibilizou nenhum Sistema de Informação (SI) que permita aos gestores públicos a simulação de cenários e análise situacional, ficando a cargo das prefeituras realizá-los. No entanto, é de conhecimento geral as dificuldades enfrentadas pelos gestores públicos municipais, sobretudo em municípios pequenos, quanto ao acesso às Tecnologias da Informação e Comunicação. Portanto, é possível afirmar que a problemática supramencionada é grave pelos seguintes motivos: i) volume de investimento na saúde, o qual é de aproximadamente 10% do Produto Interno Bruto (PIB) do país: ii) a complexidade administrativa do SUS; iii) a desassistência de pequenos municípios.

À luz de tais fatos, o presente trabalho apresenta um sistema de gestão de recursos de saúde desenvolvido sob a ótica do *Design Science Research Methodology*, com o objetivo de auxiliar os gestores públicos na análise situacional e previsão de possíveis cenários por meio de simulações de médio e longo prazo. O sistema desenvolvido foi testado e validado utilizando-se como estudos de caso municípios no Estado do Pará. Os resultados obtidos foram promissores, com os *stakeholders* satisfeitos com os relatórios gerados, destacando o potencial mercadológico latente da ferramenta proposta.

O restante deste artigo encontra-se organizado como segue. Na Seção 2 apresenta-se a metodologia adotada no presente estudo. Na Seção 3 a Portaria GM/MS 1631/2015 é resumidamente apresentada, juntamente com uma análise de competidores. Em seguida é feita a apresentação do sistema desenvolvido na Seção 4 e discutem-se os resultados na Seção 5. Por fim, as considerações finais são apresentadas na sexta e última Seção.

**2. Metodologia**

O gatilho para o desenvolvimento deste trabalho foi uma visita técnica à Central Estadual de Regulação do Estado do Pará (CER/PA), realizada por pesquisadores ligados ao Laboratório de Inteligência Computacional da Universidade Federal do Pará e ao Laboratório de Computação Aplicada da Universidade Federal do Oeste do Pará. Durante a visita, especialistas apresentaram diversas dificuldades enfrentadas pelo CER/PA e órgãos correlatos, como Secretarias Municipais de Saúde, dentre as quais, a ausência de uma ferramenta automatizada para análise situacional e previsão de possíveis cenários por meio de simulações de médio e longo prazo sob a ótica da Portaria GM/MS 1631/2015 foi a mais proeminente.

Após a primeira entrevista, os pesquisadores reuniram-se a fim de propor uma solução factível, e durante as discussões, percebeu-se que a *Design Science Research Methodology* (DSRM) *é* a metodologia que se adequa precisamente ao que viria a ser desenvolvido.

A DSRM é uma metodologia de pesquisa bastante difundida quando o objetivo final do projeto é o desenvolvimento de um produto, metodologias, estratégias e serviços, sobretudo na área de Sistemas da Informação. O DSRM é um modelo de processo composto de seis etapas postas em sequência, cujas definições e funções encontram-se descritas a seguir:

**1. Descrição do problema e motivação.** Aqui define-se o problema de pesquisa a ser abordado e justifica-se o valor da solução vislumbrada. Esta etapa serve para atomizar o modelo conceitual do problema, assim a solução poderá efetivamente capturar a sua complexidade.

**2. Definição dos objetivos para a solução.** Nesta etapa os pesquisadores devem inferir os objetivos de uma solução a partir da definição do problema e do conhecimento do que é possível e viável. Como resultado desta etapa espera-se a descrição de como o novo artefato irá lidar com problemas até então não considerados - este detalhe é imprescindível a fim de garantir a inovação no desenvolvimento do produto e, ao mesmo passo, garantir a viabilidade de execução do projeto.

**3. Projeto e desenvolvimento.** Aqui desenvolve-se e implementa-se o produto, modelo, método ou estratégia. Conceitualmente, um artefato produzido sob a DSRM pode ser qualquer objeto o qual uma contribuição de pesquisa está nele embutido. Logo, depreende-se que o artefato não é um mero produto, mas possui caráter inovador. Em particular, quando se trata de um artefato produzido no contexto de sistemas de informação, este usualmente possui também apelo mercadológico, por isso a necessidade de se consultar tanto o estado da arte quanto o estado da prática nas etapas anteriores.

**4. Demonstração.** Nesta fase, demonstra-se o uso do artefato na resolução de uma ou mais instâncias do problema. Isto envolve seu uso em experimentos e simulações; e é a partir desta etapa que o planejamento amostral, o plano de tabulação de dados, as formas de coleta e processamento de dados e instrumentos de coleta se tornam notórios.

**5. Avaliação.** A observação e mensuração do desempenho do aparelho na resolução do problema é o foco desta etapa. Esta atividade envolve a comparação dos objetivos da solução com os resultados observados na fase de demonstração. Ao final desta etapa é possível decidir sobre a necessidade de retornar à uma das fases anteriores a fim de aumentar a eficácia do artefato ou se ele continua para a próxima etapa.

**6. Comunicação.** Esta fase visa dar publicidade ao problema e sua importância, ao artefato, sua utilidade e inovação, o rigor de seu projeto e efetividade de uso. Os principais documentos construídos nesta etapa são artigos científicos, relatórios técnicos, documentos de registro de software, pedidos de patente e entrevistas.

Com a definição da metodologia a ser empregada no estudo, os pesquisadores dividiram os trabalhos em duas fases. A primeira consistiu na análise da Portaria GM/MS 1631/2015 e de softwares utilizados pelos especialistas do CER/PA. A segunda fase foi a instanciação dos elementos do DSRM, desde a identificação do problema e motivação - com os subsídios alcançados na fase 1, até a escrita deste artigo, representando a etapa de comunicação. Na próxima Seção os resultados da primeira fase da pesquisa serão apresentados.

## 3. Fundamentação teórica

### 3.1. A PORTARIA GM/MS 1631/2015

Não é possível discorrer sobre a Portaria GM/MS 1631/2015 sem antes mencionar suas predecessoras: a Portaria do Ministério da Previdência e Assistência Social (MPAS) 3046/1982 e a Portaria GM/MS 1101/2002, uma vez que estas serviram de fundamento e alicerce para a produção da retratada neste trabalho. Em 1980, o Brasil encontrava-se em um período considerado por muitos historiadores, a exemplo de (SANTAGADA, 1990, p. 121), como a "década perdida". Neste contexto, temos um Brasil sucumbindo em e endividamentos externos, com a inflação tendo atingido a taxa de 100% (ARIDA E RESENDE, 1985, p. 6). Tendo em vista o cenário apresentado, a Portaria MPAS 3046/1982 surgiu como uma das soluções de estagnação da conjuntura política e econômica, segundo Bomfim (2011, p. 369), o documento tinha o intuito de racionalizar as despesas com assistência médica da população previdenciária. Assim, foram formulados parâmetros visando reduzir os gastos, ajustar a oferta de serviços contratados e conveniados pelo Inamps, e eliminar ou reduzir as fraudes que ocorriam até então.

A Portaria MPAS 4036/1982 segue por duas vertentes: primeiramente apresentação de parâmetros de cálculos voltados para assistência da população (cobertura), e em segundo lugar cálculos de capacidade aceitável de recebimento de pacientes e produção de serviços (produtividade); tornando assim, mais parametrizado e metodológico o, relativamente novo, sistema de saúde brasileiro. (BOMFIM, 2011, p. 370). Ainda se tratando da Portaria MPAS 3046/1982, esta abrangia as consultas médicas de urgência e emergência, clínicas básicas (clínica médica, pediatria, ginecologia, obstetrícia e cirurgia geral), consultas médicas especializadas, consultas odontológicas, serviços complementares (patologia clínica e radiologia), internações (necessidade de internações e leitos) e capacidade de produção (número de consultas/hora) do profissional médico e do dentista. (SOUZA E LAGES, 2012, p. 175).

Em 12 de junho de 2002, a Portaria MPAS 3046/1982 foi revogada com a publicação no Diário Oficial da União (DOU) da nova Portaria GM/MS 1101/2002 que entrará em vigor, desenvolvida pelo Ministro da Saúde em exercício na época, Barjas Negri (MINISTÉRIO DA SAÚDE, 2002, p. 1). Segundo Souza e Lages (2012, p. 178) novos indicadores foram adicionados com a vinda desta Portaria, como por exemplo foram estabelecidos parâmetros de cobertura das consultas médicas de urgência especializadas, cirurgias ambulatoriais especializadas, procedimentos traumato-ortopédicos, ações especializadas em odontologia, patologia clínica, anatomopatologia e citopatologia, radiodiagnóstico, exames ultrassonográficos, diagnose, fisioterapia, terapias especializadas, próteses e órteses, anestesia, ressonância magnética, medicina nuclear in vitro, tomografia computadorizada e hemoterapia, entre outros.

Com base nos relatos acima, percebe-se uma ampla atualização realizada nos parâmetros de cálculo, com o acréscimo de novos indicadores. Isto fez-se necessário conforme o notório avanço e o desenvolvimento natural da nação neste período. Uma vez que esses cálculos conjecturam mais adequadamente a tendência da saúde e tendo sempre em vista a adjacência aos ODM, postula-se o oferecimento de uma melhor qualidade na gestão dos recursos de saúde.

A Portaria GM/MS 1631/2015, principal tema deste artigo, foi criada por Ademar Arthur Chioro dos Reis, então Ministro da Saúde, em 02 de outubro de 2015 - data oficial de publicação no DOU. Segundo o Diário Oficial da União (2015, p. 4), tal portaria tem o objetivo de:

> "*Definir aproximações às necessidades de saúde da população pensadas independente de restrições financeiras, séries históricas da oferta de serviços ou outros condicionantes. Rompe-se, assim, com a lógica restritiva e de controle que permeou a elaboração dos parâmetros de programação no país, desde a Portaria MPAS N° 3046, de 20 de julho de 1982, que teve grande influência nos primórdios do SUS, até a Portaria GM n° 1101, de 12 de junho de 2002."* (CHIORO, 2015, p. 4)

A Portaria GM/MS 1631/2015, apresenta também grandes diferenças em relação à sua predecessora principalmente no que tange a aferimentos mais específicos, ao cenário de consultas e doenças, entre outros. Mudanças de normativas e suplantação de Portarias fazem-se sempre necessárias, uma vez que é irremissível e extensa a mutabilidade dos serviços ligados à saúde. Por exemplo, a fim de uma explicação fundamentada em acontecimento hodierno e segundo Guerra et al. (2005, p. 228) pode-se considerar que o câncer é um importante problema de saúde pública em países desenvolvidos e em desenvolvimento, sendo responsável por mais de seis milhões de óbitos a cada ano, representando cerca de 12% de todas as causas de morte no mundo. Diante disto, buscou-se na base de dados do DATASUS informações sobre a mortalidade de câncer e conforme ilustrado no gráfico presente na Figura 1, pode-se observar uma crescente taxa de mortalidade entre 1982 e 2014.

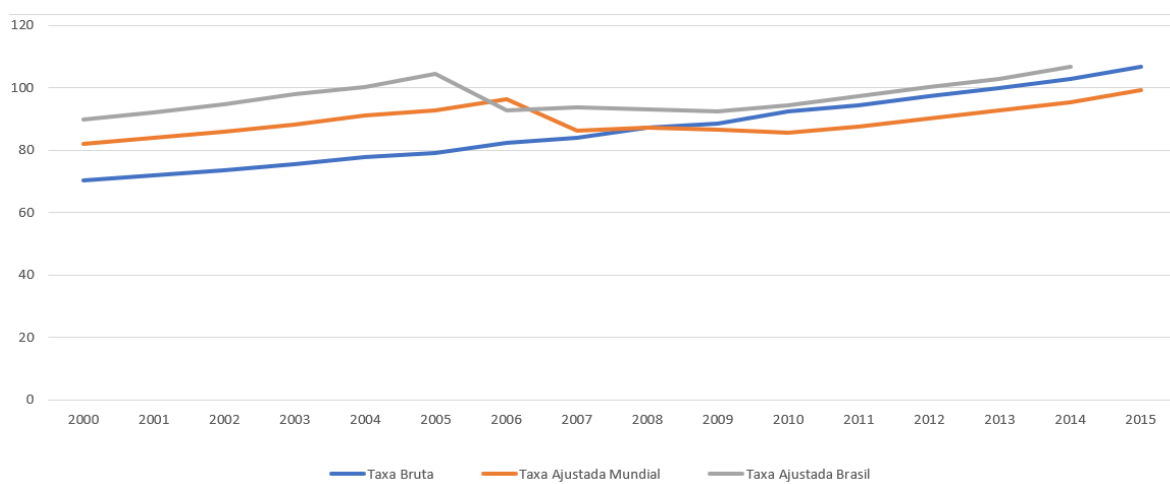

FIGURA 1 - Taxas de mortalidade por câncer, brutas e ajustadas por idade, pelas populações mundial e brasileira, por 100.000, segundo sexo, localidade e por período selecionado (DATASUS - TABNET, 2017).

Neste contexto, utilizando como exemplo o câncer, percebe-se a necessidade de aprimoração nas normativas governamentais. Em posse de ocorrências históricas e estudo em cima destas, é possível predizer futuros problemas e formular possíveis soluções. Contudo, as resoluções nem sempre são como antevistas, perante as incontáveis possibilidades de interferências externas. À vista disso, periodicamente, faz-se necessário uma reciclagem em tais decretos.

A Portaria retratada é dividida em dois capítulos, o primeiro dispõe dos cálculos preditivos ofertados pelo governo, como objetivo principal da Portaria; e o segundo capítulo reúne normas, critérios e parâmetros que constam em políticas já regulamentadas, além de disponibilizar informações reunidas de outras Portarias que possuem alguma relação.

Retornando ao primeiro capítulo da Portaria, fonte dos cálculos implementados neste trabalho, temos que este é subdividido em seções. Cada ramificação tem sua especificidade, sendo detentor dos cálculos singulares àquele domínio, de acordo com a numeração abaixo, com alguns exemplos de cada seção:

I. Atenção à gravidez, parto e puerpério:
   A. Estimativa de gestantes;
   B. Gestantes de alto e risco habitual;
   C. Estimativas de crianças por faixa etária;
   D. Procedimentos de atenção básica e ambulatorial - consultas, dosagens de vacinas exames entre outros.
II. Atenção às pessoas com doenças crônicas não transmissíveis (DCNT):
   A. Diabetes;
   B. Hipertensão;
   C. Insuficiência cardíaca;
   D. Doença Renal;
   E. Doença Pulmonar.
III. Eventos de Relevância para a Vigilância em Saúde:
   A. Doenças Sexualmente Transmissíveis;
   B. Hepatites Virais;
   C. Hanseníase;

    D. Dengue.
  IV. Atenção a saúde bucal:
    A. Atendimentos;
    B. Atenção especializada;
    C. Atenção Ambulatorial Especializada;
    D. Consultas;
    E. Atenção Básica.
  V. Atenção especializada:
    A. Médicos;
    B. Consultas;
    C. Exames especializados.
  VI. Atenção hospitalar:
    A. Leitos;
    B. Internações.
  VII. Equipamentos:
    A. Espirômetros;
    B. Tomógrafos;
    C. Pet Scan;
    D. Ultrassom convencional.
  VIII. Hematologia e Hemoterapia:
    A. Parâmetros para coleta de sangue;
    B. Parâmetros para transfusão.

  Conforme entendimento tácito, esta é a portaria vigente para predição de gastos e gestão em saúde. Uma vez diante de uma necessidade vital ou arbitrária à submissão de projetos ao SUS é fundamental a utilização da Portaria hodierna, principalmente, como justificativa para o requerimento de verba para o mesmo.

### 3.2. ANÁLISE DE COMPETIDORES

  Esta seção apresenta um estudo crítico-comparativo de sistemas de gestão de saúde, portais de notícias e buscadores de informações situadas em distintas bases de dados. Todos os sistemas, obrigatoriamente, utilizam dados reais de saúde e têm como desígnio apresentar dados epidemiológicos, indicadores de saúde, dados estatísticos e outros, que são utilizados como base de informações para análise, cálculo e tomadas de decisão para Ministérios e órgãos competentes da área. Como mencionado anteriormente, este projeto obteve apoio de um especialista pertencente à Central Estadual de Regulação (CER/PA), e seguindo as recomendações do mesmo, seis softwares foram utilizados para levantamento do estado da prática: Sala de Apoio à Gestão Estratégica (SAGE), IBGE Webcart Beta, Departamento de Informática do SUS (DATASUS), DeepAsk, Informação e Gestão da Atenção Básica (E-GESTOR), Centro de Informações e Decisões Estratégicas Em Saúde - Conecta SUS Zilda Arns Neumann (CONECTA SUS). As próximas subseções apresentam os níveis de funcionalidade e necessidades funcionais dos softwares supracitados.

### 3.2.1 Sala De Apoio À Gestão Estratégica (SAGE)

  O SAGE é uma base de dados de livre acesso que utiliza mapas e gráficos estatísticos para disponibilizar indicadores de extrema importância, como por exemplo, dados de combate à AIDS, malária e diversas doenças, índices de saúde de gestantes, mortalidade infantil, dentre outros. Foi concebido por Alexandre Rocha Santos Padilha, em 8 de março de 2012, à época, ocupante do cargo de Ministro da Saúde, através da

Portaria GM/MS 406/2012. Neste ato administrativo ficou delegado como responsabilidade do Departamento de Monitoramento e Avaliação do SUS cumprir os objetivos presentes na prescrição do Ministério da Saúde (2012), a saber: i) obter e sistematizar dados e informações produzidos pelos órgãos do MS e entidades a ele vinculadas e por outras instituições de saúde, com vistas a contribuir para o processo decisório e para o acompanhamento das políticas públicas de saúde; ii) disponibilizar, por intermédio da internet e quaisquer outros meios definidos pela Secretaria-Executiva do Ministério da Saúde, informações e análises de caráter executivo e gerencial, com o objetivo de subsidiar a tomada de decisão, a gestão e a produção de conhecimento; iii) disponibilizar regularmente relatórios de análise situacional em saúde; e iv) gerir o Portal da SAGE e o Portal da Saúde com Mais Transparência (MINISTÉRIO DA SAÚDE, 2012, p. 1).

O SAGE apresenta ainda indicadores epidemiológicos da situação atual da saúde no Brasil, como morbidade, mortalidade, violência doméstica e sexual, e dados de gestão financeira, como lista projetos e liberação de recursos. Ademais, é de grande valia a permissibilidade de exportação destes conhecimentos nos vários formatos que o sistema disponibiliza. O SAGE permite exportar para os formatos *Comma Separated Values* (CSV), *JavaScript Object Notation* (JSON) - alimentação de outros sistemas - imagens e gráficos (fundamentação das informações), entre outros. Apesar de suas qualidades singulares, detectou-se a necessidade de cálculos mais transversais a fim de suprir, de forma mais concreta, as tomadas de decisões. Deparou-se também com um déficit de fornecimento de dados para solução de problemas e principalmente gestão de saúde, como por exemplo logística de distribuição, controle financeiro e controle de estoque de suprimentos.

### 3.2.2 IBGE Webcart Beta

Pouco é registrado, seja no âmbito acadêmico-científico ou no âmbito de transparência pública, a respeito de informações sobre o surgimento do IBGE WebCart Beta3. Diante do limitado acervo bibliográfico, o website predecessor do sistema, é uma factível fonte para descrição e definição do mesmo. Segundo sua página institucional "O IBGE disponibiliza uma ferramenta simplificada para criação de cartogramas a partir dos dados contidos no canal Cidades@" (IBGE WebCart, 2016). Neste contexto, pode-se deduzir que o WebCart foi desenvolvido depois do IBGE Cidades@, uma vez que se utiliza do projeto como provisor da base de informação.

Seguindo a análise proposta, foi compreendido que suas principais funcionalidades e contribuições para o mercado são apresentar cartograficamente dados como quantitativo de nascidos vivos, estabelecimentos de saúde e demografia populacional; além de serviços de saúde, como acervo de leitos e equipamentos, diversamente detalhados. Todavia, indicadores de outras áreas como educação, transporte, entre outros que não fazem parte do escopo deste trabalho, também são concedidos pelo projeto.

Em similaridade com o SAGE, o IBGE WebCart Beta também é de livre acesso e disponibiliza seus resultados, cartograficamente distribuídos, em formatos como CSV, *Portable Document Format* (PDF) e *Scalable Vector Graphics* (SVG). Por meio da análise do IBGE WebCart Beta3 pode-se observar insuficiência no que tange a resultados mais progressistas, alcançando resoluções mais avançadas a fim de prover alicerces mais estáveis para tomadas de decisões, criação e projeção de políticas públicas mais específicas e fundamentadas.

### 3.2.3 Departamento De Informática Do Sus (DATASUS)

O DATASUS foi criado em 16 de abril de 1991. Segundo o decreto D0100, publicado no Diário Oficial da União (DOU) "Ao Departamento de Informática do SUS compete especificar, desenvolver, implantar e operar sistemas de informações relativos às atividades finalísticas do SUS, em consonância com as diretrizes do órgão setorial" (CASA CIVIL, 1991). Após submetido a análise explanada no tópico 2.2, verificou-se que tal sistema segue uma metodologia diferente dos demais, uma vez que este é formado por um conjunto de subsistemas. Neste contexto, cada módulo do sistema é responsável por trabalhar com indicadores específicos, de maneira que cada uma das aplicações é especializada e direcionada a uma área singular de conhecimento.

Um dos subsistemas pertencentes ao DATASUS é o TABNET. Ele é responsável pelas informações demográficas, indicadores de saúde, assistência social, dentre outros. Outro subsistema da composição DATASUS é o CID 10, cuja especificidade é a catalogação de doenças, mantendo assim um código único para cada uma. Tais subsistemas foram elencados dada suas importâncias e frequência de utilização por profissionais da área, em consequência da capacidade de armazenar e disponibilizar indicadores de suma importância para quantificação de doenças. As informações apresentadas e disponibilizadas pelo DATASUS são de vital importância para auxiliar nas ementas, planos e políticas públicas de saúde, visto que as bases de informações ofertada pelo sistema são a maior e mais completa fonte de conhecimento de saúde disponível. Esta ferramenta é de livre acesso, sendo uma rica fonte de consulta para produções científicas, análises de dados e profissionais da área, porém três disfunções podem ser salientadas diante da análise aplicada.

Conforme os sistemas anteriormente estudados, este possui apenas dados quantitativos e estatísticos, sendo carente pois, de cálculos que apresentem maior segurança e confiabilidade para planejamentos estratégicos e concretos. É observado também, que apesar dos subsistemas valerem-se de uma só base de informação e serem todos partes de um sistema maior (DATASUS), estes subsistemas não possuem integração entre si. Neste contexto, vale ressaltar a dificuldade encontrada para localizar as informações contidas em sua conjuntura, revelando-se assim uma plataforma não muito amigável ao usuário, uma vez que se faz necessário muitas interações com o sistema para alcançar a exportação ou visualização dos dados. Tal argumento havia sido previamente identificado e sinalizado pelo especialista adjunto do projeto.

### 3.2.4 DeepAsk

A DeepAsk foi criada em 22 de julho de 2013. Este sistema funciona como um provedor de conhecimento de distintos temas, como saúde, economia, transporte, comunicação, entre outros. Seu real propósito é pouco difundido, porém, segundo a sua área institucional, "A missão do DeepAsk é centralizar os dados abertos da internet e torná-los mundialmente acessíveis para pesquisa e análise. Através de infográficos e mapas de acesso interativo, o DeepAsk facilita o acesso aos dados abertos para milhares de pessoas em todo o mundo" (DeepAsk, 2015).

Realizando a análise proposta com o auxílio do especialista supramencionado, pode-se observar que, diferente dos demais, este sistema tem uma premissa peculiar, dada sua forma única de apresentação dos dados. Neste cenário, infere-se que o enfoque da DeepAsk não é disponibilizar mais uma base de dados, uma vez que sua metodologia é disponibilizar a informação a partir da análise de indicadores provenientes de inúmeras bases de dados, porém no formato portal de notícias. Apesar de grande valia na sua forma de disseminar o conhecimento, tendo em vista o seu público alvo, e propiciar alicerces sólidos para produção acadêmico-científica, tal metodologia recai em uma limitada aplicabilidade, no que se refere ao oferecimento de base de informação para desenvolvimento de sistemas.

### 3.2.5 Informação E Gestão Da Atenção Básica (E-GESTOR)

Segundo o material de apresentação, produzido pela equipe do Núcleo de Tecnologia da Informação (NTI) do Departamento de Atenção Básica (DAB), "o E-gestor foi desenvolvido para centralizar e organizar o acesso aos sistemas; permitir maior agilidade na realização dos cadastros; reforçar a identidade dos sistemas ofertados pelo Departamento e reunir informações de apoio aos Gestores estaduais e municipais" (NTI - DAB, 2016). O E-GESTOR em sua Versão 6 apresenta duas áreas de acesso, sendo a primeira pública, a qual dispõe de alguns relatórios básicos no âmbito financeiro, geográfico e de saúde; e a segunda restrita, permitida apenas para gestores e técnicos, estaduais e municipais, órgãos federais e outras instituições. Embora o profissional adjunto do projeto pertença à CRE/PA, este também não tem acesso ao sistema, impossibilitando assim uma análise mais completa desta área. Ainda sobre o E-Gestor, obteve-se a informação que tal sistema é mais utilizado a nível de modelo de relatórios, ou seja, baseia-se em sua estrutura para produção de relatórios de auxílio à gestão. Neste contexto, não é conclusivo se o sistema possui serventia no âmbito de disponibilidade de base de dados. Contudo, caso este o tenha em sua área restrita, seu difícil acesso (dado os limitantes cargos permitidos), torna-o inviável para aplicação neste trabalho.

### 3.2.6 Centro De Informações E Decisões Estratégicas Em Saúde - Conecta Sus Zilda Arns Neumann (CONECTA SUS)

O projeto Conecta SUS foi criado pela Secretaria da Saúde do Estado de Goiás (SES-GO), em 3 de dezembro de 2014. Segundo a própria Secretaria "o centro de informações é uma nova forma de fazer Saúde Pública, com informações atualizadas, disponibilizadas em tempo real, que darão total apoio aos gestores nas instâncias estadual e municipal, na tomada de decisões" (SES-GO, 2014). Pela análise proposta e executada neste trabalho, o sistema produzido pelo projeto Conecta SUS foi entendido como um aglomerador e sinalizador dos dados disponibilizados pelo DATASUS.

Este sistema é muito utilizado pelos gestores como canal de localização de dados, uma vez que é isento de restrição e, como citado anteriormente, o DATASUS não possui uma interface amigável ao usuário, consequentemente, dificultando o encargo de seus utilizadores, com ênfase nos que não detêm total familiaridade com a disposição das informações. Entretanto, é de grande amparo à contribuição realizada, porém o recorrente problema encontrado nos demais sistemas analisados, dito que não é disponibilizado uma base de acesso própria, também se faz presente no mesmo.

### 3.3 CONSIDERAÇÕES DA SEÇÃO

Por fim, após completar a análise destes sistemas correlatos, pode-se concluir que ainda se faz necessária uma aplicação que utilize toda a gama de informação disponibilizada pelos sistemas elencados, a fim de aplicar cálculos mais complexos e preditivos, como os dispostos na Portaria GM/MS 1631/2015. Sendo assim, percebeu-se um déficit no que tange a predição de saúde, mais especificamente com a utilização de portarias governamentais. Neste contexto, julga-se necessário o desenvolvimento de um sistema que trabalha com tal aspecto, a fim de simplificar e agilizar o processo de tomada de decisão dos gestores da área, o qual é apresentado a seguir.

## 4. Resultados

Conforme mencionado anteriormente, a metodologia escolhida para a condução desta pesquisa foi a DSRM, apresentada na Seção anterior. As subseções a seguir apresentam a instanciação dos elementos do DSRM para o presente trabalho.

### 4.1. PROBLEMA E MOTIVAÇÃO

Instanciando os elementos do DSRM para a presente proposta, temos que o principal problema consiste na falta de um sistema da informação capaz de auxiliar os gestores na realização dos cálculos dispostos na Portaria GM/MS 1631/2015, cuja implementação é impreterível à gestão eficiente dos recursos públicos.

Este problema foi evidenciado por meio de *observação e entrevistas* com especialistas do CRE/PA. Em tal processo, identificou-se que na ausência de um sistema implementador da referida Portaria, utilizava-se planilhas que implementavam a Portaria predecessora (1101/2002 - já revogada). Em segundo lugar, temos que apenas a seção 6 da portaria atual tem seus cálculos implementados em planilhas (profissionais da área o fizeram dada a frequência de utilização); os cálculos das demais seções, quando necessárias, são realizados manualmente. Considerando a totalidade das seções, estimou-se que apenas 15% dos cálculos da Portaria estavam automatizados em planilhas eletrônicas, tornando a análise situacional e preditiva uma tarefa morosa e burocrática.

As motivações para lidar com este problema são diversas, primeiramente, a quantidade de recursos destinados à saúde é enorme - o investimento na saúde chega aos 10% do Produto Interno Bruto do País, ultrapassando a cifra de R$ 100 Bilhões em 2016. Além deste montante, a complexidade administrativa do Sistema Único de Saúde Brasileiro e a desassistência de pequenos municípios no tocante ao acesso à Tecnologias de Informação e Comunicação é notória e impacta significativamente na burocratização do sistema de saúde brasileiro.

### 4.2. OBJETIVOS

Sendo assim, definiu-se como objetivo principal deste trabalho o desenvolvimento de um sistema de informação que implemente, de forma automática, os cálculos dispostos na Portaria GM/MS 1631/2015, e que auxiliasse os gestores públicos na análise situacional e previsão de possíveis cenários por meio de simulações de médio e longo prazo.

Para atingi-lo, os seguintes objetivos específicos foram traçados:
1. Analisar as principais ferramentas disponíveis na área, identificando suas vertentes de trabalho, tecnologias utilizadas, plataformas de disponibilização, funcionalidades e necessidades ainda não implementadas
2. Compreender e descrever o propósito da Portaria GM/MS 1631/2015, bem como seu histórico de mudança
3. Investigar a melhor forma de distribuição do software proposto
4. Escolher tecnologias para o desenvolvimento do sistema dados os requisitos
5. Implementar o sistema de informação proposto
6. Validar o sistema desenvolvido

Os objetivos específicos 1 e 2 foram feitos na primeira fase do Projeto, já os objetivos 3-6 pertencem à segunda fase do projeto e são descritas nas Subseções a seguir.

**4.3. PROJETO E DESENVOLVIMENTO**

O projeto e desenvolvimento do sistema proposto sofreu influência de diversos fatores, sendo a estratégia de distribuição do software a variável que impactou diretamente nas definições arquiteturais e tecnológicas. Frente ao exposto, nesta seção são apresentadas a estratégia de distribuição de software, o ciclo de desenvolvimento adotado, as especificações técnicas (arquitetura e tecnologias utilizadas) e, por fim, as funcionalidades do sistema.

**4.3.1 Ciclo de Desenvolvimento e distribuição do software**

A DSRM não está relacionada com nenhum processo de desenvolvimento de software específico. No entanto, devido a própria organização da Portaria GM/MS 1631/2015 em capítulos, optou-se pela abordagem iterativa incremental, com implementação das seções de cada capítulo em cada ciclo, incrementando o software a cada iteração com a implementação das demais seções. A cada término de capítulo, eram realizadas validações com conjunto de dados já calculados anteriormente por um especialista. Convém destacar que foram utilizadas algumas características de metodologias ágeis devido a equipe reduzida.

Concomitantemente com a definição das diretrizes do processo de desenvolvimento, foram discutidas as formas de distribuição do software proposto durante as entrevistas e observações dos trabalhos dos especialistas, atendendo o objetivo específico 3. Tal especificação se fez imperativa no início do processo de desenvolvimento pois ela seria fator determinante dos requisitos funcionais e não funcionais do software. Nesse contexto, optou-se por realizar o modelo *Software as a Service*. Segundo Cancian e Thinkstrategies (2009, p. 14), define-se SaaS como:

> *"Uma solução de software que fica hospedada no provedor do serviço e está disponível na web. Este software é oferecido como um serviço, e é acessado pelos usuários através da Internet, sem a necessidade de implantar e manter uma infraestrutura de TI. O cliente possui direitos sobre seus dados e uso do software, mas em nenhum momento precisa adquirir uma licença ou comprar o software como se fosse um produto.* (CANCIAN; THINKSTRATEGIES, 2009, p. 14)"

Em vista de tais dados, entende-se o SaaS como uma distribuição do produto, sem licença de uso, como de costume em softwares locais, mas sim com licença de acesso ao mesmo, uma vez que este não precisará ser instalado e não necessitará de uma infraestrutura prévia no que tange ao usuário, tornando assim, como um conjunto, menos custoso ao público alvo. Uma das grandes vantagens do modelo é a escalabilidade do mesmo, ou seja, o usuário é onerado apenas com o que ele utiliza. Nesse contexto o usuário normalmente realiza o pagamento apenas do que utiliza e caso necessite de uma demanda maior pode realizar um upgrade em seu plano/pacote para consumir o necessário.

Diante deste modelo acima citado, deparou-se com duas formas de oferta do software: a primeira através de uma mensalidade, dessa forma o utilizador obteria acesso ilimitado às buscas do sistema. Porém, conforme adendo do especialista, a utilização da portaria dar-se-á para submissões de projetos ou para justificativa de ingresso no programa SUS, nesse caso, alguns usuários a utilizam de forma esporádica, sendo, portanto, um custo, relativamente, alto para os mesmos. Os cálculos também podem ser utilizados para preparação de relatórios de secretarias e coordenadorias governamentais - neste caso, a utilização do mesmo será mais alta. Ademais ter-se-ia um alto grau de dificuldade em mensurar o investimento, uma vez que a variação na utilização, por perfil de usuário, seria muito alta.

A segunda forma de oferta do software é o pagamento através de cada busca realizada, dessa forma o usuário não declina a um pagamento desnecessário, uma vez que só é cobrado se for realizada uma busca. Neste contexto, por um lado depara-se com uma maior facilidade de alcançar o valor adequado, posto que não haverá pagamento desnecessário por falta de uso do sistema, porém por outro, este método pode ser custoso às secretarias e coordenadorias, uma vez que sua utilização será mais frequente.

Optou-se, portanto, por se desenvolver, inicialmente, no segundo modelo - pagamento por busca. Uma vez tudo implementado, o software sendo utilizado e munido da base de buscas realizadas por cada usuário, pode-se aplicar um estudo a fim de alterar a forma de pagamento, construindo pacotes de busca por quantidade mensal, homogeneizando os valores, no que se refere à quantidade de utilização de cada perfil. Com esta definição, construiu-se o principal artefato desta fase, o fluxo do *Business Process Modeling* apresentado na Figura 2.

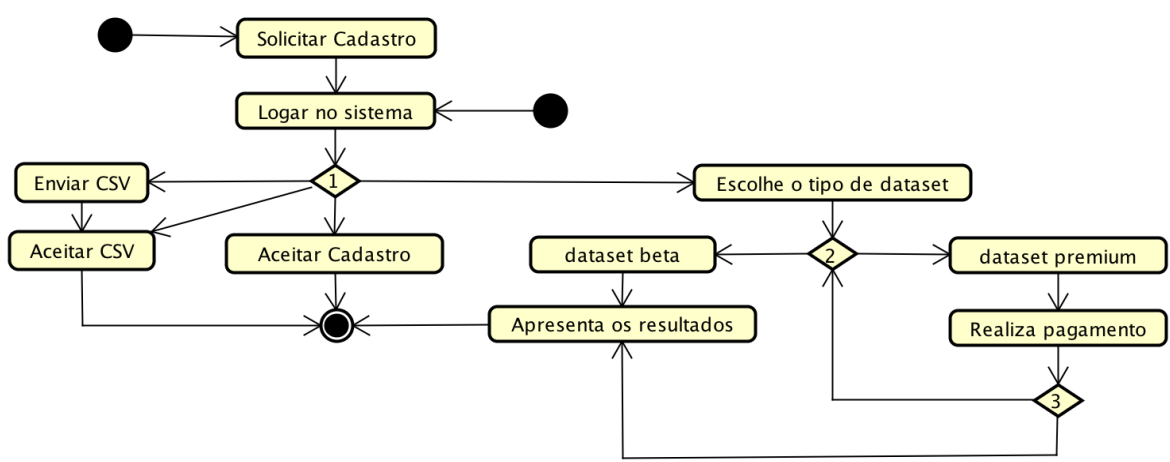

FIGURA 2 - Diagrama de atividades do SI proposto.

Após o *login*, tem-se quatro possíveis funções do sistema a serem seguidas, representada pelo número "1" na imagem. O primeiro RF, mais à esquerda (superior) da imagem, é o envio de bases de dados para o sistema. O segundo RF, mais à esquerda (inferior) da imagem, é a aceitação das bases de dados. O terceiro RF, ao centro da imagem, é a aceitação de novos cadastros. Por fim, o quarto RF, mais à direita da imagem, são as buscas ou solicitações de cálculos. Ademais, as buscas do sistema se sub-ramificam em duas possibilidades, representado pelo número "2" na imagem: a primeira é a busca beta e a segunda é a busca *premium*. Faz-se jus ressaltar que no sistema proposto existem três tipos de usuários. O usuário administrador, de uso exclusivo dos mantenedores do sistema, porquanto, do desenvolvedor. O usuário beta, realizador apenas de cálculos do tipo beta. O usuário *premium*, compreende aqueles que em um período de trinta dias já realizaram ao menos uma busca *premium*.

**4.3.2 Especificações técnicas**

Com o modelo de negócio definido, refinaram-se os requisitos funcionais e não funcionais, dentre os quais, merecem destaque para as definições da: i) plataforma, ii) arquitetura, iii) tecnologias, iv) entrada e saída. Acerca do primeiro item, os especialistas do CER/PA indicaram que a plataforma Web seria a mais indicada, uma vez que a maior parte dos softwares utilizados na área residem nessa plataforma. Ademais, segundo um dos especialistas, a portabilidade de sistemas web, uma vez que estes podem ser acessados por qualquer dispositivo seria um atrativo a mais.

Considerando as características da plataforma e o ciclo de desenvolvimento escolhidos, optou-se por adotar o modelo *model-view-controller*, onde as camadas podem ser implementadas de forma modular. Segundo Pires, Nascimento e Salgado (2012, p. 527):

> *"O padrão MVC (Model-View-Controller) sugere uma arquitetura de software dividida em componentes, viabilizando com clareza o desenvolvimento de um código organizado e enxuto, e posteriormente, a reciclagem e manutenção do sistema sem dificuldade e com segurança. Porém, a independência dos componentes só será atingida se houver uma organização do sistema em camadas para garantir a escalabilidade, eficiência e a reusabilidade. "* (PIRES; NASCIMENTO; SALGADO, 2012, p. 527)

Em posse das definições acima descritas, partiu-se para a escolha das tecnologias, tais quais: linguagem de programação, sistema de gerenciamento de banco de dados (SGBD), interface e *framework*. A linguagem escolhida foi Python, pois esta fornece suporte à Programação Web e também à análise de dados, necessário para as análises preditivas. O SGBD escolhido foi o MySQL, dado a sua robustez e simplicidade de utilização. Os *frameworks* arquitetural e de *front-end* escolhidos foram, respectivamente, o *DJango* e o *Twitter Bootstrap*, devido ao suporte da comunidade de desenvolvimento e aderência aos requisitos não funcionais estabelecidos.

**4.3.3 Funcionalidades**

O processo de elicitação das funcionalidades culminou na construção do diagrama de casos de uso, onde percebem-se quatro atores e as funcionalidades que eles podem acessar, conforme apresentado na Figura 3.

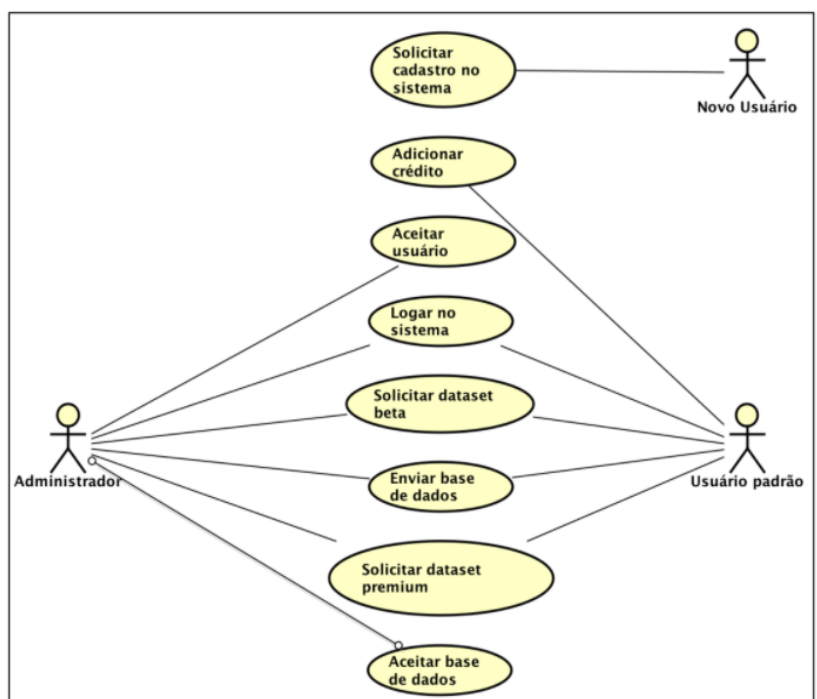

FIGURA 3 - Modelo de casos de uso construído.

O Administrador é o responsável por gerenciar todo o sistema, controlando os novos usuários e submissão das bases de dados. Optou-se pelo papel do administrador a fim de reduzir a complexidade da implementação de camadas de segurança. Por hora, a análise manual é mais interessante para evitar a exploração de eventuais falhas do sistema. São atividades exclusivas do administrador a aceitação de novos usuários e a avaliação das bases de dados. A Figura 4 apresenta a tela inicial de um usuário administrador.

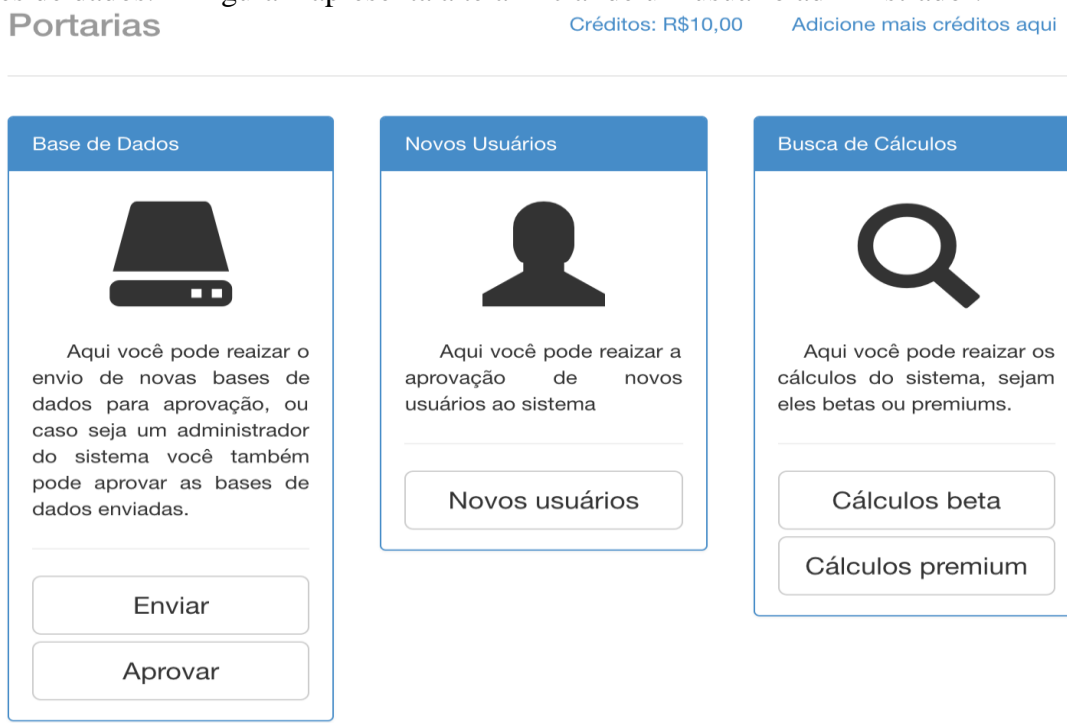

FIGURA 4 - Tela inicial para o usuário Administrador.

As funcionalidades descritas nas Figuras 3 e 4 encontram-se resumidas a seguir:

- **Novo usuário:** solicitação de cadastro.
    - Este usuário é caracterizado ao realizar uma solicitação de cadastro. Neste contexto, tem-se que este é um tipo de usuário de "aguardo" a ingresso no sistema.
- **Usuário beta:** logar no sistema, solicitar dataset beta.
    - Este usuário é um usuário de teste ou demonstração do sistema, como dito anteriormente. Nesse contexto, este não usufrui de todos os benefícios do software.
- **Usuário premium:** logar no sistema, solicitar dataset beta, solicitar dataset premium, enviar base de dados.
    - Este usuário é um usuário comum do sistema, com todos os benefícios do mesmo e com acesso à maior parte das funcionalidades.
- **Administrador:** logar no sistema, solicitar dataset beta, solicitar dataset premium, enviar base de dados, aceitar base de dados, aceitar cadastro.
    - Este usuário é o administrador da rede, sendo o mantenedor do sistema e/ou o desenvolvedor do mesmo.

Após carregar os datasets, os usuários devem filtrar os indicadores carregados para receber os relatórios. Conforme explicado anteriormente, o sistema permite ao usuário realizar dois tipos de buscas, a busca do tipo beta e a busca do tipo premium, que necessitará de um pagamento. Ambas as buscas serão redirecionadas para a mesma página de filtro, que é apresentada na Figura 5.

FIGURA 5 - Página de busca do sistema.

Na página de filtro o usuário deve selecionar primeiramente o Estado que deseja realizar a busca e se deseja fazer esta busca por um município ou por uma região de saúde do Estado selecionado. Em conformidade à escolha anterior, o sistema apresentará as regiões de saúde ou os municípios disponíveis daquele Estado, respectivamente. Por fim o usuário deve selecionar o ano em que deseja aplicar os cálculos da Portaria. Para a Agência Nacional de Saúde Suplementar (ANS, 2017), pode-se definir região de saúde como:

"*É um espaço geográfico contínuo constituído por agrupamentos de municípios limítrofes, delimitado a partir de identidades culturais, econômicas e sociais e de redes de comunicação e infraestrutura de transportes compartilhados, com a finalidade de integrar a organização, o planejamento e a execução de ações e serviços de saúde.*"(ANS, 2017)

Neste contexto, tem-se que uma região de saúde é o conjunto de alguns municípios de um Estado. Ademais, caso o usuário selecione uma região de saúde, será disponibilizado para ele todos os anos disponíveis para esta. Tendo em mente o conceito de região de saúde, tem-se que a disponibilidade dos anos é feita desde o mais antigo até o mais recente, dos municípios que fazem parte desta região, ou seja, ainda que se tenha apenas um município da região com registro em um determinado ano, este ainda será mostrado no sistema. Caso o usuário tenha escolhido um município, será apresentado para o mesmo os anos disponíveis para aquele município. A apresentação dos resultados é melhor discutida na próxima seção, de demonstração da ferramenta.

**4.4. DEMONSTRAÇÃO**

A demonstração do artefato produzido ocorreu, tal como previsto na metodologia de desenvolvimento, em etapas iterativas e incrementais. Para fins de validação, utilizaram-se dados fictícios onde os cálculos já haviam sido realizados manualmente pela equipe do CER/PA.

| cod_estado | sigla_estado | estado | cod_municipio | municipio | cod_regiao | ano regiao | 2010 populacao | sinasc (nv) | 2011 populacao | sinasc (nv) |
|---|---|---|---|---|---|---|---|---|---|---|
| 15 | PA | Pará | 150380 | Jacunda | 15004 | Lago de Tucurui | 10000 | 1000 | 20000 | 2000 |

FIGURA 6 - Exemplo do dataset de entrada.

O dataset contém o código do Estado, a sigla, nome, além do código do município, seu nome e o código da região de saúde a qual ele pertence. Também contém os anos e os dois principais dados demográficos de entrada, o tamanho da população e o número de nascidos vivos. Ao final de cada interação, os resultados eram comparados e ajustes no código eram realizados. Ao término da implementação das seções da portaria e validação, demonstrou-se que o sistema desenvolvido resolvia o problema dos cálculos manuais para a análise situacional. A análise preditiva, por ser análoga, também seria resolvida. Em posse do sistema validado, partiu-se para a avaliação do mesmo, a qual é apresentada na próxima seção.

**4.5. AVALIAÇÃO**

O software foi avaliado por meio de dois estudos de caso, utilizando-se dados dos municípios de Água Azul do Norte e Ananindeua, ambos no Pará. O primeiro estudo de caso destinou-se a avaliar a análise situacional, e para tal, utilizaram-se dados do ano de 2015 de Água Azul do Norte, localizado no Sudeste Paraense, pois os cálculos também já haviam sido realizados manualmente pela equipe do CER/PA. Em 2015, o município de Água Azul do Norte possuía 26.305 habitantes e teve 187 nascidos vivos.

O Quadro 1 apresenta alguns indicadores calculados a partir da Portaria vigente. A análise situacional pode ser feita medindo-se os números postos pelos cálculos e os números reais observados. É importante ressaltar que a Portaria é válida para todo o território nacional e é de conhecimento geral que as estatísticas são significativamente diferentes nas diversas regiões.

QUADRO 1 - População de referência para as internações por tipo de leito.

| Cálculo | Máximo anual |
|---|---|
| Obstetrícia | 196 |
| Neonatologia | 187 |
| Pediatria clínica (Menores de 15 anos: 24,2%) | 6.366 |
| Pediatria cirúrgica (Menores de 15 anos: 24,2%) | 6.366 |
| Clínica - 15 a 59 anos (População de 15 a 59 anos: 64,8%) | 17.046 |
| Clínica - 60 anos e mais (População 60 anos e mais: 11%) | 2.894 |
| Cirurgia - 15 a 59 anos (População de 15 a 59 anos: 64,8%) | 17.046 |
| Cirurgia – 60 anos e mais (População 60 anos e mais: 11%) | 2.894 |

Por meio da análise do Quadro 1 e respeitando as diferenças sócio econômicas dos municípios interioranos, é possível observar que para o município de Água Azul do Norte o número de leitos de obstetrícia estimado pela Portaria é 196, enquanto para neonatologia o número é menor, 187. Isto ocorre levando-se em consideração o índice médio de mortalidade infantil e natalidade média. Contudo, tratando-se de uma cidade interiorana e com indicadores econômicos abaixo de média, o cenário mais provável é uma quantidade maior de obstetrícia e menor de neonatologia. Tais interpretações foram corroboradas pelos especialistas consultados.

O município de Ananindeua foi utilizado no segundo estudo de caso. Para avaliar as análises situacionais e preditivas foram usados os dados de 2016 e a previsão de 2020. Em 2016, Ananindeua - segunda maior cidade do Estado do Pará, possuía 510.834 habitantes e teve 8.974 nascidos vivos. Para 2020, utilizando dados preditos pelo IBGE, Ananindeua deverá ter 521.675 habitantes e 9.893 nascidos vivos. O Quadro 2 mostra o resultado para os gastos previstos e realizados com cardiologia em 2016.

QUADRO 2 - Resumo dos cálculos com os valores para cardiologia em Ananindeua no ano de 2016.

| | Cardiologia | | | | |
|---|---|---|---|---|---|
| Código | Cálculo | Máximo anual | Média Mensal | Valor Unitário | Gasto |
| | Quantidade de médicos 40 horas semanais- Cardiologista | 33 | - | - | - |
| 03.01.01.007-2 | Consulta Médica Cardiologia | 30.650 | 2554 | R$ 10,00 | R$ 25.541,70 |
| 02.11.02.004-4 | Holter | 1.533 | 128 | R$ 30,00 | R$ 3.831,26 |
| 02.05.01.003-2 | Ecocardiografia Transtoracica | 8.173 | 681 | R$ 39,94 | R$ 27.203,61 |
| 02.11.02.006-0 | Teste ergométrico | 3.065 | 255 | R$ 30,00 | R$ 7.662,51 |
| 02.05.01.002-4 | Ecocardiografia Transesofágica | 102 | 9 | R$ 165,00 | R$ 1.404,79 |
| 02.05.01.001-6 | Ecocardiografia de estresse | 102 | 9 | R$ 165,00 | R$ 1.404,79 |
| 02.08.01.002-5 | Cintilografia miocárdica em situação de estresse | 1.022 | 85 | R$ 408,52 | R$ 34.780,98 |
| 02.08.01.003-3 | Cintilografia miocárdica em situação de repouso | 1.022 | 85 | R$ 383,07 | R$ 32.614,20 |
| 02.08.01.008-4 | Ventriculografia radioisotópica | 5 | 0 | R$ 176,72 | R$ 75,23 |
| 02.11.02.001-0 | Cateterismo cardíaco | 2.043 | 170 | R$ 614,72 | R$ 104.673,29 |
| 02.11.02.002-8 | Cateterismo cardíaco em pediatria | 5 | 0 | R$ 653,72 | R$ 278,29 |

A primeira coluna representa o código de cada item, o qual já está preenchido no banco de dados carregado pelo administrador. O mesmo vale para os valores, que são atualizados periodicamente, conhecidos por Tabela SUS. Por exemplo, a cada consulta médica de cardiologia é pago pelo SUS o valor de R$ 10,00. Para calcular o custo mensal de cada item, multiplica-se a média de atendimentos/procedimentos por mês pelo valor. Por exemplo, para cateterismo cardíaco, a Secretaria de Saúde de Ananindeua pode gastar na média mensal o valor de R$ 104.673,29.

Ao final da apresentação dos resultados aos *stakeholders*, a facilidade de utilização e o nível de satisfação quanto ao uso foi avaliado de forma qualitativa em entrevista aberta. Pelo número reduzido de especialistas consultados, análises quantitativas tornavam-se inviáveis. O *feedback* recebido é que a ferramenta possui potencial mercadológico, precisando de aperfeiçoamentos quanto à sua interface e análises suplementares.

### 4.6. COMUNICAÇÃO

A comunicação da produção do artefato descrito nas subseções anteriores ocorreu pela defesa do trabalho de conclusão de curso do primeiro autor; na apresentação do sistema como caso de sucesso pelo professor orientador; na prestação de serviços de consultoria à três gestores públicos que não permitiram sua identificação e na construção deste artigo.

## 5. Discussões

Um dos pontos discutidos com os especialistas do CER/PA foi de o sistema receber como entrada também os dados reais, além dos dados preditos pois, em posse desses valores, os gestores públicos podem comparar com o que foi gasto - automatizando a análise situacional. No entanto, os *stakeholders* consultados descartaram esta funcionalidade por dois motivos: o primeiro é relacionado à quantidade de informação e o preenchimento manual, que é suscetível à erros; o segundo, é relacionado ao sistema de repasse de verba que é difícil de ser explicado.

Percebeu-se também uma falta de interesse por parte dos *stakeholders* entrevistados, já que os mesmos possuíam também empresas de consultoria na área, prestando serviço de análise situacional. Acerca da análise preditiva, ela é baseada nas duas estatísticas de entrada do sistema, população atual e nascido vivos. O IBGE divulga as projeções para os próximos anos, permitindo que o sistema também disponibilize análise preditiva.

## 6. Considerações finais

O Sistema Único de Saúde brasileiro é uma referência internacional na área, servindo de exemplo para outros países que buscam sistemas de saúde com maior qualidade e equidade. Considerando a dimensionalidade continental do Brasil e da complexidade no sistema administrativo do SUS, é possível concluir que a gestão da distribuição de recursos é uma tarefa não-trivial. A normativa corrente que versa sobre a gerência de recursos para saúde, a portaria GM/MS 1631/2015, dispõem sobre todos os produtos e serviços disponibilizados.

Apesar da abrangência e relevância, ressalta-se que o Ministério da Saúde não disponibilizou nenhum Sistema de Informação que permita aos gestores públicos a simulação de cenários e análise situacional, ficando a cargo das prefeituras realizá-los. É de conhecimento geral as dificuldades enfrentadas pelos gestores públicos municipais, sobretudo em municípios pequenos, quanto ao acesso às Tecnologias da Informação e Comunicação. Portanto, é possível afirmar que a problemática supramencionada é grave considerando volume de investimento na saúde, a complexidade administrativa do SUS e, sobretudo, a desassistência de pequenos municípios.

À luz de tais fatos, o presente trabalho apresentou um sistema de gestão de recursos de saúde desenvolvido sob a ótica do *Design Science Research*, com o objetivo de auxiliar os gestores públicos na análise situacional e previsão de possíveis cenários por meio de simulações de médio e longo prazo. Para obter estas informações, primeiramente, os usuários devem carregar os datasets com os parâmetros necessários e, posteriormente, devem filtrar os indicadores carregados para receber os relatórios. Neste filtro dos indicadores, usuário deve selecionar primeiramente o Estado que deseja realizar a busca e se deseja fazer esta busca por um município ou por uma região de saúde do Estado selecionado. A partir da escolha anterior, o sistema apresentará as regiões de saúde ou os municípios disponíveis daquele Estado. Por fim o usuário deve selecionar o ano em que deseja aplicar os cálculos da Portaria. Para ter esse acesso, o software será oferecido no modelo SaaS, ou seja, o usuário realizará o pagamento apenas do que utilizar e, caso necessite de uma demanda maior, pode realizar um upgrade em seu plano/pacote para consumir o necessário.

A validação do sistema, se deu através de dois estudos de caso, ambientados em municípios do interior do estado do Pará. Os resultados são promissores, uma vez que em ambos foi possível simular diversos cenários para médio e longo prazo. Considerando esses resultados, para trabalhos futuros, devem ser realizados mais estudos de caso, incluindo um número maior de municípios de diferentes portes. Além disso, uma pesquisa quantitativa, deve ser realizada para avaliar a satisfação dos gestores públicos em relação ao serviço prestado e deste modo, melhor mensurar o potencial mercadológico da ferramenta.